\documentclass[pre,aps,twocolumn,showpacs,floatfix]{revtex4}
\usepackage{graphics}
\usepackage{graphicx}
\usepackage{epsf}

\begin{document}

\title{Fast Monte Carlo simulations and singularities in
the probability distributions of non-equilibrium systems}

\author{A. Bandrivskyy$^1$, S. Beri$^1$, D.~G. Luchinsky$^1$, R. Mannella$^2$
and P.~V.~E. McClintock$^1$}

\affiliation{$^1$Department of Physics, Lancaster University,
Lancaster
LA1 4YB, UK\\
$^2$Dipartimento di Fisica and INFM, Universit\`a di Pisa, Italy}

\date{\today}
\widetext

\begin{abstract}
A numerical technique is introduced that
reduces exponentially the time required for Monte Carlo
simulations of non-equilibrium systems. Results for the
quasi-stationary probability distribution in two model systems are
compared with the asymptotically exact theory in the limit of
extremely small noise intensity. Singularities of the
non-equilibrium distributions are revealed by the simulations.
\end{abstract}

 \maketitle
\newcommand{\ud}{\mathrm{d}}

The understanding of fluctuations in systems away from thermal
equilibrium is a problem of long standing in statistical physics
\cite{Onsager:53}. Well known examples of physical phenomena in
which non-equilibrium fluctuations play a particularly important
role include e.g.\ switching of polarization in lasers
\cite{Keay:95}, switching between different configurations in
proteins \cite{Serpersu:83}, the transition to instability in
Josephson junctions \cite{Kautz:96}, and chemical reactions
\cite{Dykman:99a}.

In non-equilibrium systems, where symmetries of detailed balance
are broken, no general methods exist for the calculation of even
basic quantities like the probability distribution. This is a case
where numerical and asymptotic theoretical methods for
investigating the probability distribution are of particular
importance.

Theoretical approaches, such as WKB-like or path-integral methods,
are available in the limit of small noise intensity, $D\to 0$
\cite{Freidlin:70b, McKane:89, Dykman:90a, Dykman:99a}. In
particular the theory suggests that a solution to the problem of
non-equilibrium fluctuations requires an understanding of the
dynamics of deviations from the steady state \cite{Onsager:53} and
an analysis of singularities in the non-equilibrium potential
\cite{Graham:84,Smelyanskiy:97}. Some ideas about how to extend
the existing ($D\rightarrow0$ limit) theory for still small but
finite noise intensities have recently been suggested
\cite{Dykman:99b, Hanggi:00a, Bandrivskyy:02}.

The main numerical technique used to verify theoretical
predictions, and to analyse the behavior of the dynamical system
under study, is Monte Carlo simulation. The theory gives an
asymptotically exact solution in the $D \to 0$ limit. In contrast,
$D$ in the numerical simulations is necessarily finite. Typically,
the time required for Monte Carlo simulations grows exponentially
as $D \to 0$. This meant that theoretical predictions of
interesting singular structures, and of the non-equilibrium
probability distribution \cite{Graham:84,Jauslin:87}, for long
remained untested either experimentally or by numerical
simulation. Moreover there was no clear understanding of how the
picture changes for small but still finite noise intensities.

Approaches that have been tried to speed up the simulations have
focused mainly on finding optimal fluctuational paths and rates of
transition between stable states of a system (e.g.\ efficient
transition path sampling \cite{Dellago:98} and dynamics importance
sampling \cite{Woolf:98}, following the earlier suggestion of
\cite{Pratt:86}). In \cite{Crooks:01} the path sampling method was
adapted for non-equilibrium systems. Based on the same idea, the
umbrella sampling technique was suggested to estimate the
probability of reaching any point in the phase space of an
equilibrium system starting from a fixed initial state
\cite{Dellago:98}. A technique for improving sampling in
equilibrium systems by splitting up the probability packets was
introduced in \cite{Huber:96}. So far, however, no general
algorithm has been suggested, able to give both the whole
probability distribution and dynamical information like the
optimal fluctuational paths for small noise intensities for
non-equilibrium systems.

In this Letter we introduce a numerical method that enables the
time required for Monte Carlo simulations to be reduced by an
exponentially large factor. It is applicable to generic
two-dimensional non-equilibrium systems, does not require any {\it
a priori} knowledge about the system apart from its dynamical
equations of motion, and it allows the quasi-stationary
probability distribution to be computed directly over the whole
phase space. Using this method, we reveal for the first time
singular behavior of the non-equilibrium distribution in numerical
simulations, and we show that the results are in quantitative
agreement with the asymptotic theory.

The central idea is to perform the simulations in sequential
steps. We construct the quasi-stationary distribution, patching
together intermediate results: we start from one of the steady
states and gradually move away from it. We find that the time
required for the simulations at each step is reduced by an
exponentially large factor as compared to the standard technique:
if the time necessary for a conventional Monte Carlo simulation
technique is $T$, our modified method would require only time $T_m
\approx N T \exp^{-(N-1)\frac{\Delta \Phi}{D}}$, where $N$ is the
number of steps involved and $\frac{\Delta \Phi}{D}$ is their
separation in terms of the logarithm of the probability
\footnote{It is illuminating to compare simulations of the Duffing
system by the fast and conventional techniques: the results of
Fig.\ 3 took us $\sim$15 minutes to simulate with $D=0.02$,
whereas for the same noise intensity it takes $\sim$4 days of
standard Monte Carlo simulation to obtain comparable statistics
close to the boundary of attraction.}.

We first explain the method on a very simple equilibrium
stochastic system, and then we apply it to two much-studied
non-equilibrium systems and compare the numerical results with
theoretical predictions. To illustrate the technique, we consider
an overdamped Brownian particle moving in a bistable Duffing
potential $U(x)=-x^2/2+x^4/4$
\begin{eqnarray}
\label{Brown} \dot x = -U'(x) +\xi(t),
\end{eqnarray}
where $\xi(t)$ is zero-mean white Gaussian noise with intensity
$D$ and moments
\begin{eqnarray}
\label{corr} \left<\xi(t)\right> = 0, \quad
\left<\xi(t)\xi(0)\right> =2D\delta (t). \nonumber
\end{eqnarray}
The probability distribution is completely defined by the
potential $U(x)$, and is of the Boltzmann form $\rho(x) \propto
\exp(-U(x)/D).$ As in the case of a non-equilibrium system (where
the probability distribution is not defined by a potential) a
standard Monte Carlo technique can be used to deduce $\rho(x)$.
Numerical integration \cite{rm:00} of the Langevin equation
(\ref{Brown}), assuming the system to be located initially at one
of the potential minima $x_m$, gives the discrete probability
distribution $\rho(x)$, peaked at $x_m.$ The potential can be
deduced as $\Phi(x) \propto - D \ln \rho(x)$. If the noise
intensity is very small, the system fluctuates in a close vicinity
of $x_m$ and large deviations from it are extremely rare.
Accordingly, the conventional Monte Carlo technique cannot be used
to study the dynamics of optimal escape paths, or the properties
of the probability distribution far from the potential minima: for
small noise intensities the statistics required cannot in practice
be collected within a realistic time.

In order to overcome this problem, we start from the distribution
already obtained near $x_m.$ We fix two probability levels
$\rho_i$ and $\rho_f$, lying well within the region where the
numerical $\rho$ is accurate, with $\rho_f<\rho_i$ corresponding
to two levels in the potential $\Phi_i$ and $\Phi_f$, and two
coordinates $x_i$ and $x_f$, as shown in Fig.\ 1. We require the
levels $\rho_i$ and $\rho_f$ to be fairly different, such that the
corresponding $x_i$ and $x_f$ are sufficiently separated: the
distance between them must exceed $\sqrt{D h}$, where $h$ is the
integration time step used in the Monte Carlo simulation, and must
also exceed the discretization step $\Delta x$ in the discrete
probability distribution.

The next step of the simulation is started from the level $\Phi_i$
(putting the system at $x=x_i$ as its initial condition). If the
system starts to evolve along a fluctuational trajectory (towards
the boundary of attraction) we just follow its natural dynamics
according to (\ref{Brown}) and collect the statistics for building
the probability distribution in the usual way. If the system
starts with a relaxation trajectory (towards $x_m$), or when it
crosses the boundary $x_i$ due to relaxation some time later, we
stop the simulation and reinject the system back to the initial
state $x_i.$ In this way we simulate the full dynamics of the
system at higher levels of the potential $\Phi(x)>\Phi_i$ (in the
region of coordinate space $x>x_i$ for this particular case).
Thus, in the subsequent simulation step we follow only those
fluctuations that have already attained a certain level in the
potential $\Phi_i$, without waiting for this exponentially slow
event to happen. In this way, a new piece of the probability
distribution is built with a time saving $\sim \exp{\Phi_i/D}$
compared to a simulation starting from the potential minimum
$x_m$. The computed new piece of the potential $\Phi_2(x)$ is
shown as the upper curve in Fig. 1.

\begin{figure}[ht]
\label{Fig_1}
\includegraphics[scale=0.44]{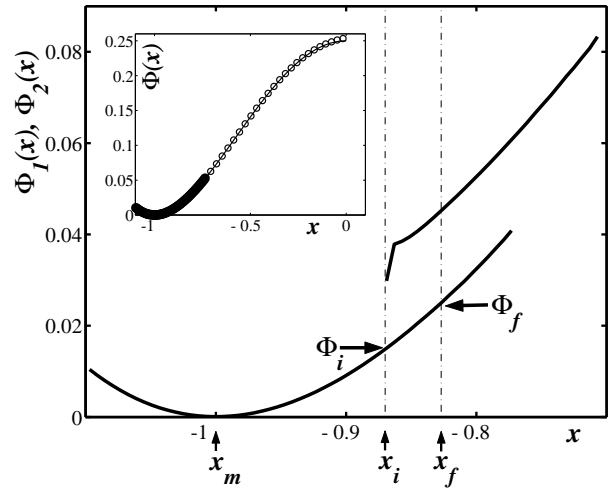}
\caption{The first ($\Phi_1(x)$, lower curve) and second
($\Phi_2(x)$, upper curve) pieces of the inferred potential
$\Phi(x)$ for the system (\ref{Brown}) with $D=0.005$. The
discontinuity in the gradient of $\Phi_2(x)$ near $x_i$ is an
artefact due to a boundary effect in the calculation of the
discrete probability distribution. To avoid this problem
$\Phi_1(x)$ and $\Phi_2(x)$ are merged at the point $x_f$ and the
initial part of $\Phi_2(x)$ is discarded. We normalize $\Phi_1(x)$
choosing $\Phi_1(x_m)=0$, and each successive piece of $\Phi(x)$
is normalized in order to match with the previous one at the point
where they join. Inset: the inferred potential $\Phi(x)$ for the system
(\ref{Brown}) with $D=0.005$. The new technique (circles) is
compared with standard Monte Carlo simulations (bold line) and
with the Duffing potential $U(x)$ (thin line). }
\end{figure}

The two pieces of the inferred potential (the original $\Phi_1(x)$
and the new $\Phi_2(x)$) are then merged at $x_f$ by a simple
shift. Continuing this procedure, the probability distribution and
the corresponding potential can be built, step by step, for the
whole region of interest. The inset in Fig. 1 shows the resultant
potential, built from 13 such pieces between the minimum at
$x_m=-1$ and the maximum at $x=0$. It coincides closely with the
Duffing potential $U(x)$ itself. The potential $\Phi(x)$ is thus
inferred within a region of coordinate space that is inaccessible
in a conventional simulation (shown as bold curve for comparison).
We stress that no {\it a priori} knowledge of the dynamics has
been used in the simulations, and that the method is robust to
choice of parameters.

In the case of a two dimensional system, the procedure remains
essentially the same. The main difference is that, instead of
identifying two points $x_i$ and $x_f$, we need to identify two
closed lines of constant probability. One line is a boundary line
for starting simulations from, and the other is a reference line
for matching together different pieces of the probability
distribution (see Fig.2 for clarification) \footnote{In the case
of an $N$-dimensional system, this would correspond to
$(N-1)$-dimensional surfaces of constant probability dividing the
phase space into separate regions.}. The crucial point of our
technique is that, in starting the simulations from the boundary
line, we must not perturb the natural dynamics of the system. This
implies that we should consider the reinjection location
probability (RLP) along the boundary line corresponding to
$\rho_i$. Starting from the second step of the simulations, the
system should be reinjected back according to the RLP after it
relaxes across the boundary. We emphasize that the RLP is not the
same as the probability distribution $\rho(\bf x)$, which is
constant on the boundary line. The RLP is an additional important
measure which describes local discrete dynamics in the
neighborhood of the boundary line. It is a distribution along the
boundary of how often the system crosses it.

In an equilibrium system, detailed balance provides a symmetry
that can be used to reinject the system back at the boundary
level, without any need to compute the RLP. For non-equilibrium
systems, however, this procedure is inapplicable. The RLP should
be considered separately (and calculated explicitly) for the
particular system being investigated. It can be obtained from an
analysis of the finite difference equation corresponding to the
model. In the limit of small integration time step the probability
to cross the boundary is proportional to the diffusion-related
term in the finite difference equation. Then the RLP is simply
proportional to the projection of the vector orthogonal to the
boundary onto the coordinate affected by the noise $\xi$. It can
also be computed numerically.

For non-equilibrium systems, the limit of small noise intensity is
of particular importance. A sufficiently small $D$ gives rise to
the possibility of revealing the non-equilibrium potential
\begin{eqnarray}
\label{NP} \Phi({\bf x})=\lim_{D\rightarrow0} -D \ln \rho({\bf
x}), \nonumber
\end{eqnarray}
directly through a numerical experiment. Observations of the
predicted singular shape of $\ln \rho({\bf x})$, and of its
dependence on $D$, are thus of considerable interest.

\begin{figure}[ht]
\label{Fig_2}
\includegraphics[scale=0.44]{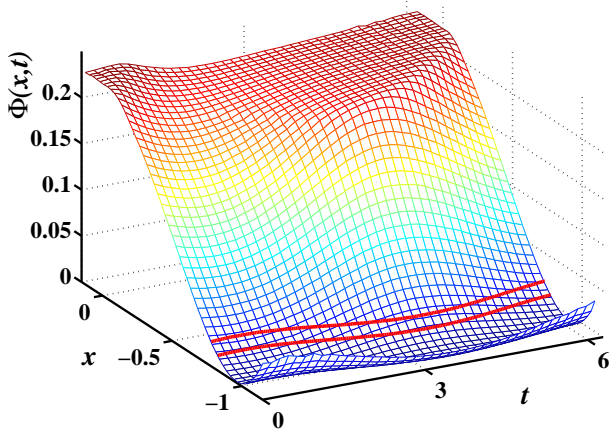}
\caption{The whole inferred $\Phi(x,t)$ for the system
(\ref{Brown_per}) for $A=0.1$, $\Omega=1$, $D=0.005$. Two lines
are the lines of constant probability found after the first step
of simulations. The corresponding levels of probability were
chosen as $\Phi_i=3 D$ and $\Phi_f=5 D$.}
\end{figure}

\begin{figure}[ht]
\label{Fig_3}
\includegraphics[scale=0.44]{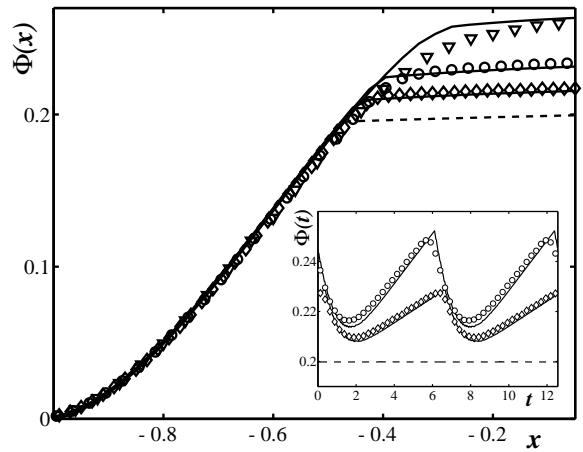}
\caption{A time section of the inferred $\Phi(x,t=4.1)$ for the
system (\ref{Brown_per}) with $A=0.1$, $\Omega=1$, and different
noise intensities: $D=0.005$ (diamonds); $D=0.01$ (circles); and
$D=0.02$ (triangles). The theoretical predictions are shown by
full lines for finite noise intensities, and by dashed line for $D=0$.
Inset: oscillations of $\Phi(x,t)$ at the boundary of attraction
for different noise intensities.}
\end{figure}

As a first example of a nonequilibrium system, consider the
periodically-driven overdamped Duffing oscillator
\begin{eqnarray}
\label{Brown_per} \dot x = -U'(x) + A\cos{\Omega t} + \xi(t).
\end{eqnarray}
We infer $\Phi(x,t)$ as $-D \ln \rho(x,t)$. This quantity
corresponds to the theoretical ``global minimum of the modified
action" in the Hamiltonian theory of large fluctuations
\cite{Bandrivskyy:02} and, in the limit $D\rightarrow0$, it
becomes the non-equilibrium potential.

The complete $\Phi(x,t)$, constructed from 12 such pieces, is
shown in Fig.~2 and a time section of $\Phi(x,t)$ calculated for
different noise intensities together with the results of
theoretical calculations (Hamiltonian theory including the
prefactor) \cite{Bandrivskyy:02} is shown in Fig.~3. The RLP in
the simulations can be taken as constant if a small enough
integration time step is used in the scheme. A small difference
between the theory and the simulations results appears for larger
noise intensities then the asymptotic theory starts to break down.

\begin{figure}[t]
\label{Fig_4}
\includegraphics[scale=0.44]{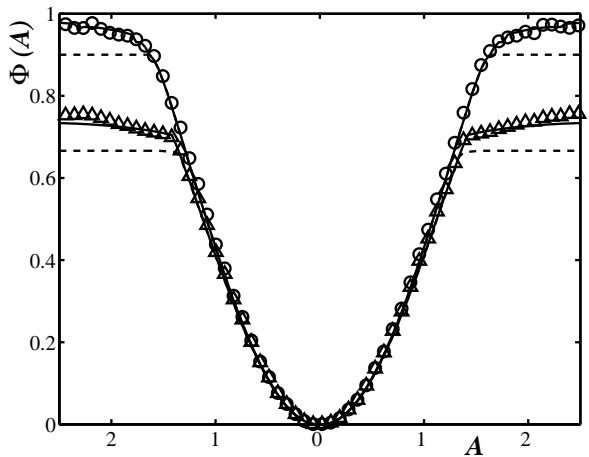}
\caption{A section $(x=y)$ of the inferred $\Phi(A)$ for the
system (\ref{Van}) with $\omega_0=1$, noise intensity $D=0.01$ and
$\eta=0.25$ (circles); and $\eta=0.5$ (diamonds). Theoretical
predictions are shown in each case for $D=0$ (dashed curves) and
$D=0.01$ (full curves).}
\end{figure}

As a second, more complicated, nonequilibrium example, consider
the inverted Van-der-Pol oscillator
\begin{equation}
\label{Van}
 \ddot x + 2 \eta (1-x^2) \dot x + \omega_0^2 x = \xi(t)
\end{equation}
Here, in order to be able to merge more easily the different
pieces of $\Phi(x,y)$, we apply a coordinate transformation from
$x$ and $y=\dot x$ to amplitude $A$ and phase $\phi$ ($x=A
\cos(\phi), \ y=-A \omega_0 \sin(\phi)$). The probability
$\rho(x,y)$ can be then analyzed in the $(A, \phi)$ coordinate
space. This makes the problem very similar to the periodically
driven Duffing oscillator: the only difference is the RLP which,
in the case of the Van der Pol oscillator, turns out to be
strongly modulated. It is essential for this modulation to be
taken into account when reinjecting the system back to the
boundary of constant probability. Two sections of $\Phi(x,y)$,
obtained from the simulations for different parameters $\eta$, are
compared with the theory in Fig.~4. Again, the agreement between
numerics and theory is excellent.

The non-equilibrium systems considered in this Letter share the
same structure of singularities. Using the fast Monte Carlo
simulations we reveal plateaus, the essentially flat regions in
the probability distribution, which can be observed close to
boundaries of attraction. They result from a purely dynamical
effect that is not associated with the flatness of any potential.
We have shown that its origin is related to switching between
different types of optimal fluctuational path, and it is a general
feature of non-equilibrium systems with metastable states
\cite{Bandrivskyy:02,Bandrivskyy:02pre}. The switching lines
\cite{Smelyanskiy:97}
 are revealed as lines along which the
``global minimum of the modified action" $\Phi(\bf x)$ exhibits
sharp bends -- corresponding to the predicted line at which the
non-equilibrium potential is non-differentiable. In the boundary
region we found the oscillations of the probability distribution
and their dependence on noise intensity (see the inset in Fig.3)
discussed in the recent publications
\cite{Dykman:99b,Hanggi:00a,Maier:01a}. Using the simulations we
demonstrated noise induced shift of the singularities and the
optimal escape path, which has stimulated a new step in the
development of the theory \cite{Bandrivskyy:02}.

We emphasize that the singularities can be confidently observed
only in the limit of extremely small noise intensity, and
therefore that the use of our new technique is crucial in that it
reduces by an exponentially large factor the time required for
Monte Carlo simulations. In addition to being fast, it preserves
dynamical information, can be modified to analyse optimal
fluctuational paths, is applicable to the energy-optimal control
problem \cite{Khovanov:00}, and can be further extended to
encompass higher dimensional systems and maps.

The work was supported by the Engineering and Physical Sciences
Research Council (UK), the Joy Welch Trust (UK), the Russian
Foundation for Fundamental Science, and INTAS.

\end{document}